\documentclass[aps,twocolumn,showpacs,superscriptaddress]{revtex4-1}
\usepackage{amsmath,amssymb}
\usepackage{graphics,graphicx}
\usepackage{dcolumn,bm}
\usepackage{psfrag}
\newcommand{\cu}
{\affiliation{Department of Physics, University of Calcutta,
92 Acharya Prafulla Chandra Road, Kolkata 700009, India.}}

\begin{document}

\title{Frozen states and active-absorbing  phase transitions   of the  Ising model on networks}

\author{Abdul Khaleque}%
\cu
\author{Parongama Sen}%
\cu

\begin{abstract}
A zero temperature quench of the Ising model  is known to lead to a frozen steady
state on random and small world networks. 
We study such quenches on  random scale free networks (RSF) and compare the scenario
with that in the Barab\'{a}si-Albert network (BA) and the  Watts Strogatz (WS) addition type network.
While frozen states are present in all the cases, the RSF shows
an order-disorder phase transition of mean field nature as in the WS model as well as the existence of two absorbing phases separated by an active phase.
The WS network also shows an active-absorbing (A-A) phase transition occurring at the known order-disorder transition point.
The comparison of the RSF and the BA 
network results show interesting difference in finite size dependence.

\end{abstract}


\maketitle

\section{Introduction}
\label{sec1}

Dynamics on networks is a topic on which extensive works have been
done in recent years. Phenomena which have been studied include evolution of spin systems, 
opinion dynamics, disease spreading dynamics, etc. \cite{Barrat,psen-book}. 
The dynamical picture is  quite different from that on regular lattices due to 
the topological features of network. For example, one can define the dynamics in different ways for the voter model on a network while these 
rules become identical on lattices \cite{caste}.

The study of Ising model
on networks has revealed a number of interesting features  when 
static properties are considered.  
Even in one dimension, when randomly new links are added (or existing links rewired) as in a Watts Strogatz (WS) network \cite{WS},
 one gets a phase transition \cite{Weigt,Gitterman,Niko} which occurs
with mean field criticality \cite{Kim,Herrero,Hong}. On Euclidean networks, indications of both mean field type and finite dimensional-like phase transitions have been 
shown to exist by varying the relevant parameter \cite{achat-psen}. 
On scale free networks \cite{Leone,Golt,Igloi}, the transition temperature shows a logarithmic
increase with the system size which is perhaps the most surprising result \cite{Alek,Bian,Herrero1,Viana}.

While considering ordering dynamics on regular lattices at zero temperature for the Ising model using Glauber dynamics, 
it is known that for any dimension greater than one, freezing occurs with a probability dependent on the dimension \cite{redner}. This happens 
when one considers a completely random initial condition. On random graphs and 
networks, one encounters similar freezing phenomena  which depend on the 
density of added links \cite{svenson,hagg,boyer,pratap}.

On random networks or graphs,  the
 evolution of the Ising model from a completely random state
shows that the system does not order.
A freezing effect was observed   and  although there could
be an emergent majority of nodes with either spin up or spin down state, domains of nodes with opposing spins
survive \cite{svenson,hagg}. 
Careful observations show  that  the disordered state is not an absorbing state \cite{castellano}. 
It is instead a stationary active state,
with some spins flipping, while keeping the energy constant. The number of
domains remaining in the system is just two. The qualitative picture is then the same as on
regular lattices for $d > 2$ \cite{redner},  the system wanders forever in an iso-energy set of states.
The distribution of  the steady state residual energy (which is identical to the number of bonds between oppositely oriented spins apart from a constant) 
for the Ising model on a random networks
was investigated \cite{baek}.
It was found that the distribution typically shows two peaks, one very close to the actual ground state where the residual
energy is zero and one far away from it.

Dynamics of the Ising model on the WS network  with restricted rewiring has also been considered.
 Here  initially
a spin is connected to its four nearest neighbours
 and then only the second nearest neighbour links are rewired  with probability $p$.
The system therefore always remains connected.
Under the  zero temperature Glauber dynamics,
freezing effect was observed
for any $p\neq0$ \cite{biswas}.% ??.

In this paper, we have considered in detail the variation of the relevant thermodynamic quantities as 
functions of time for the zero temperature dynamics of Ising model
on various networks. Our main emphasis is on the Ising model on  scale free networks at temperature $T=0$ as such studies 
have not been made  so  far to the best of our knowledge. Apart from the question whether
the equilibrium state is reached or not, we have also explored the nature of the state in case it does not.
We are  interested to see whether 
any active-absorbing phase transition occurs as the system parameters are
varied. 
Both the random scale free network (RSF) and the Barab\'{a}si-Albert (BA) network have been considered for the study.
Although many results are known for the WS network, we have explored  specifically the 
possibility of  an active-absorbing phase transition in this network.
The BA model is studied to make direct comparison with the random scale free network, where results can be  quite different \cite{albert1}.
Also, comparison with respect to
issues like freezing and absorbing phase transition 
may be made for the RSF and the WS networks.

In section II, we describe the network models and the dynamical evolution.
The quantities which have been calculated are defined in Section III. The results are presented in the next section and in the last section we summarise and 
discuss the studies made. 

\section{THE NETWORK MODELS and DYNAMICS}
We have considered 
 three different types of network: (a) Random scale free, (b) Barab\'{a}si-Albert and (3)
 Watts Strogatz (addition type) \cite{WS} network. We describe in brief how these networks are generated
and the dynamical evolution process. In this section we also include a brief discussion of how the numerical data had been analysed in previous studies.

\subsection{Random scale free (RSF) network} In the random scale free network the degree distribution follows a power law
but otherwise the network is random.
To generate random scale free network \cite{albert1,doro,boguna,newman,albert,psen}, we assigned the degree of  each node using
the power law:
\begin{equation}
{ \mathcal P}(k) \sim k^{-\gamma}, 
\end{equation}
where $k$ is the degree of node and $\gamma$ is the characteristic degree exponent. The minimum value of   $k$ is $1$ and 
the maximum cut-off value is $\surd N$, where $N$ number of nodes.
This cut-off value ensures that there is no correlation \cite{cutoff}. 
%Here the total number of links is always even ??.
Starting from the node with the maximum degree, links have been established with   randomly selected  distinct nodes.
%We have taken care of the false connections.
%Degree distribution of this network follows power law, which indicates that degree  distribution has no characteristic scale.

\subsection{ Barab\'{a}si-Albert (BA) network} 
Barab\'{a}si-Albert network is a growing network where new nodes are joined to existing nodes
with preferential attachment. We start with the three fully connected nodes.
Subsequently,   a single node is  added at  a time to the network which is linked to 
one existing node. The probability that  the new  node is  connected
to the existing $i$th node with degree $k_i$ is given by \cite{albert,albert1},  
\begin{equation}
 \Pi (k_i) = k_i/ {\sum_{j} {k_j}}.
\end{equation}
Degree distribution of the network is a power law with exponent $\gamma=3$;
%\begin{equation}
$ P(k) \sim k^{-3}$ \cite{albert1}.
%\end{equation}

\subsection{Watts Strogatz (addition type) model} Addition type WS network is a one  dimensional regular
chain with two nearest neighbour links as well as with some extra randomly  connected long range links.
Here the long range links have been added with probability $q/N$ (total long range links $\sim O(N)$), where $N$ is the number of  nodes and $q$ is a parameter, 
which denotes the number of extra long range links per node on an average. So the average degree per node of this network is $2+q$,
which is a finite quantity as $q/N \to 0$ in the thermodynamic limit. It is known \cite{Gitterman} that an order-disorder transition occurs at $q=1$.

\subsection{Dynamics of Ising model on networks}
\label{ising-net}
The Hamiltonian of the  Ising system in these networks can be expressed as
\begin{equation}
H= -{\sum_{i<j} {J_{ij}S_iS_j}}\\,
\end{equation}
where  $S_i = \pm1$ and $J_{ij} = 1$ when sites $i$ and $j$ are connected and zero otherwise.
Starting with the random configuration, single spin flip energy minimizing Glauber dynamics has been used
to update the spin. In this dynamics a randomly selected spin is flipped if the energy of the updated configuration is lowered.
It is  flipped with probability $1/2$ if the energy remains unchanged on flipping. 
Fifty different network configurations have been considered and for each network 
hundred different initial configurations have been taken. The results are averaged over these configurations. We have considered system sizes $N$ upto $1600$.
Periodic boundary 
condition has been used for the WS model which is embedded  in real space. 

Previous studies  have shown that for the thermally driven phase transition, not only mean field critical behaviour exists for the Ising model
 on small world networks but finite size scaling is also valid there \cite{Kim,Herrero,Hong,achat-psen}.
The data collapse was obtained by rescaling the data and the scaling argument occurring in the scaling function was found
to be of the familiar form $\epsilon N^{1/ \tilde{\nu}}$ (with $\tilde{\nu}=2$) where $\epsilon$ denotes the deviation from the critical point.
% The scaling function occurring in the finite size scaling is of
%the form $tN^{1/ \tilde{\nu}}$ where $t$ denotes the deviation from the critical point, and 
$\tilde{\nu}$ is argued to be equal to $\nu d$ where $\nu$ is the correlation length exponent and $d$  the effective
dimension of the system. With mean field critical exponent $\nu=0.5$,
 $d$ turns out to be equal to the upper critical dimension; $d=4$.
% However dimension for network is usually
%not uniquely defined. In the Random scale free (RSF) network we also get phase transitions by tuning the parameters and 
%finite size scaling is also seem to be valid. Hence we take the effective  dimension to be equal to $4$ following the earlier studies.

\section{Quantities calculated}
We have estimated the following quantities in the present work.

1. Magnetisation: $m(t)$ has been calculated by taking the average of the absolute values of the magnetisation, $m={\langle| \sum_{i}S_i|\rangle}/{N}$, as a function of time.
Since evolution to both up spin dominated and down spin dominated configurations are possible, the absolute
value is taken to compute the configuration average. Distribution of saturation value of magnetisation   has also been estimated for random scale free (RSF) network.

2. Residual energy: $E_r (t) = E(t)-E_g$ where $E_g$ is the equilibrium energy of the ground state per spin and $E(t)$ the energy per spin at time $t$.
Residual energy measurement indicates the closeness to the equilibrium ground state.
Since we are employing a zero temperature quench, the equilibrium ground state configuration corresponds to either all spins up or down.

3. $P_{flip}(t) = N_{flip}(t)/N$, where $N_{flip}$ is the number of spin flips at time $t$, has been studied as a function of time.
 We count all the
spin flips, i.e., if the same spin flips more than one time, all these occurrences are taken into account.

4. Freezing probability $F$ is defined as the probability that the system does not reach the true ground state.
It has  been calculated as a function of the relevant parameters.
For $T=0$, the known ground state is the ferromagnetic state with $m(t\rightarrow \infty)=1$.
 A frozen configuration will have an absolute value of magnetisation less than $1$.

5. Whether or not the dynamics continue, the system always reaches an iso-energy state. Time  $\tau$ to reach the iso-energy 
 state  has been calculated as a function of the relevant parameter for random scale free (RSF) network.

The saturation  values have been denoted using appropriate subscript, e.g. $m_{sat}$ for magnetisation.
%Any configuration freezes (does not reach the equilibrium ground state) with this probability.

\section{Results}
\subsection{Random scale free network}
\begin{figure}[!h]
\includegraphics[width=5cm,angle=270]{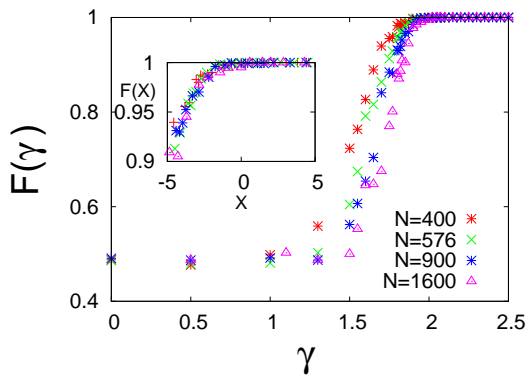}
\caption{(Color online) RSF: Variation of freezing probability $F(\gamma)$
 as a function of $\gamma$ for different system size $N$.
Inset shows the data collapse where $F(\gamma)$ has been plotted against $X=(\gamma-\gamma_c^f)N^{1/{\tilde\nu}}$.
 \label{sat_frz_rsf}
}
\end{figure}
\begin{figure}[!h]
\resizebox{88mm}{!}{\includegraphics {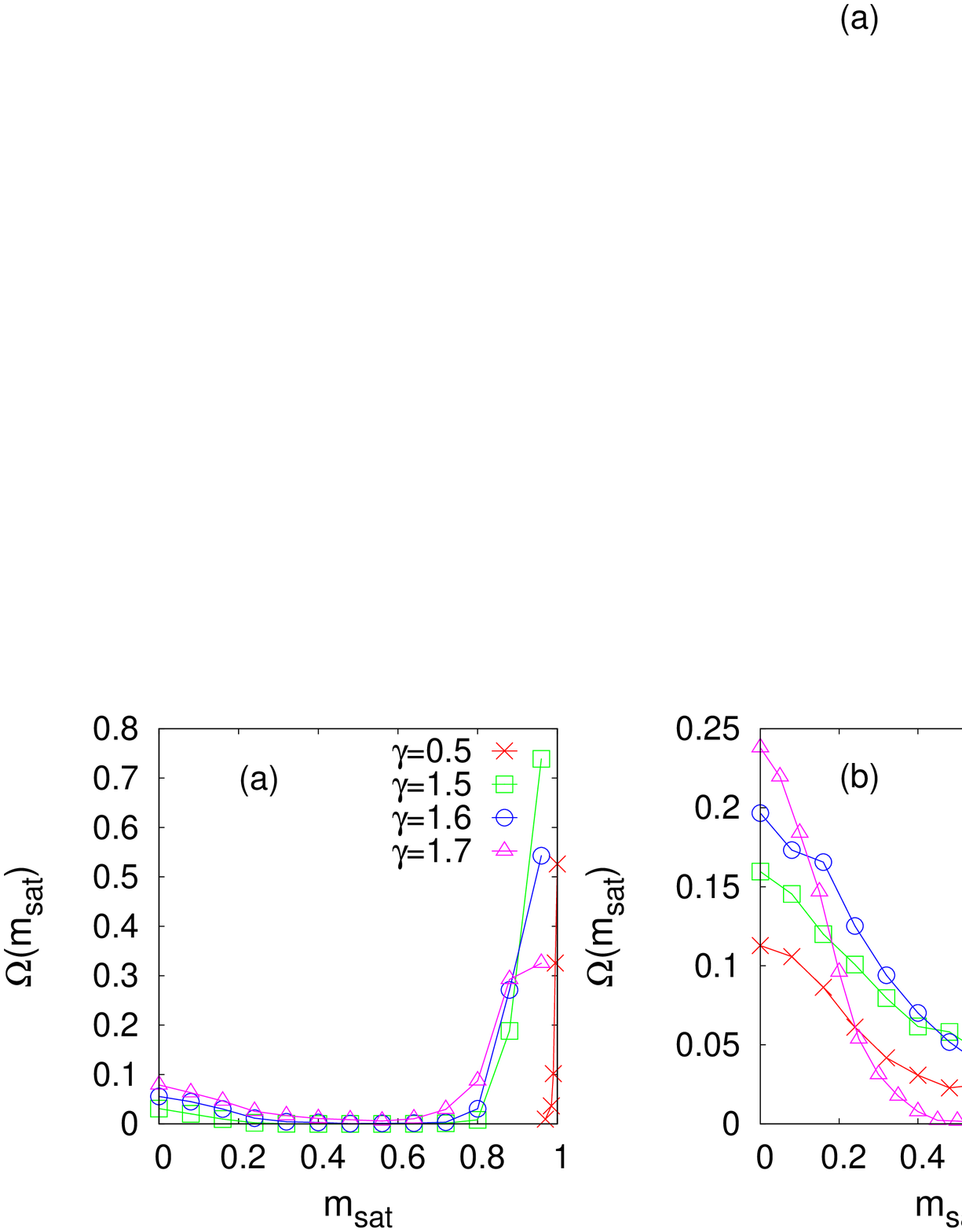}}
\resizebox{88mm}{!}{\includegraphics {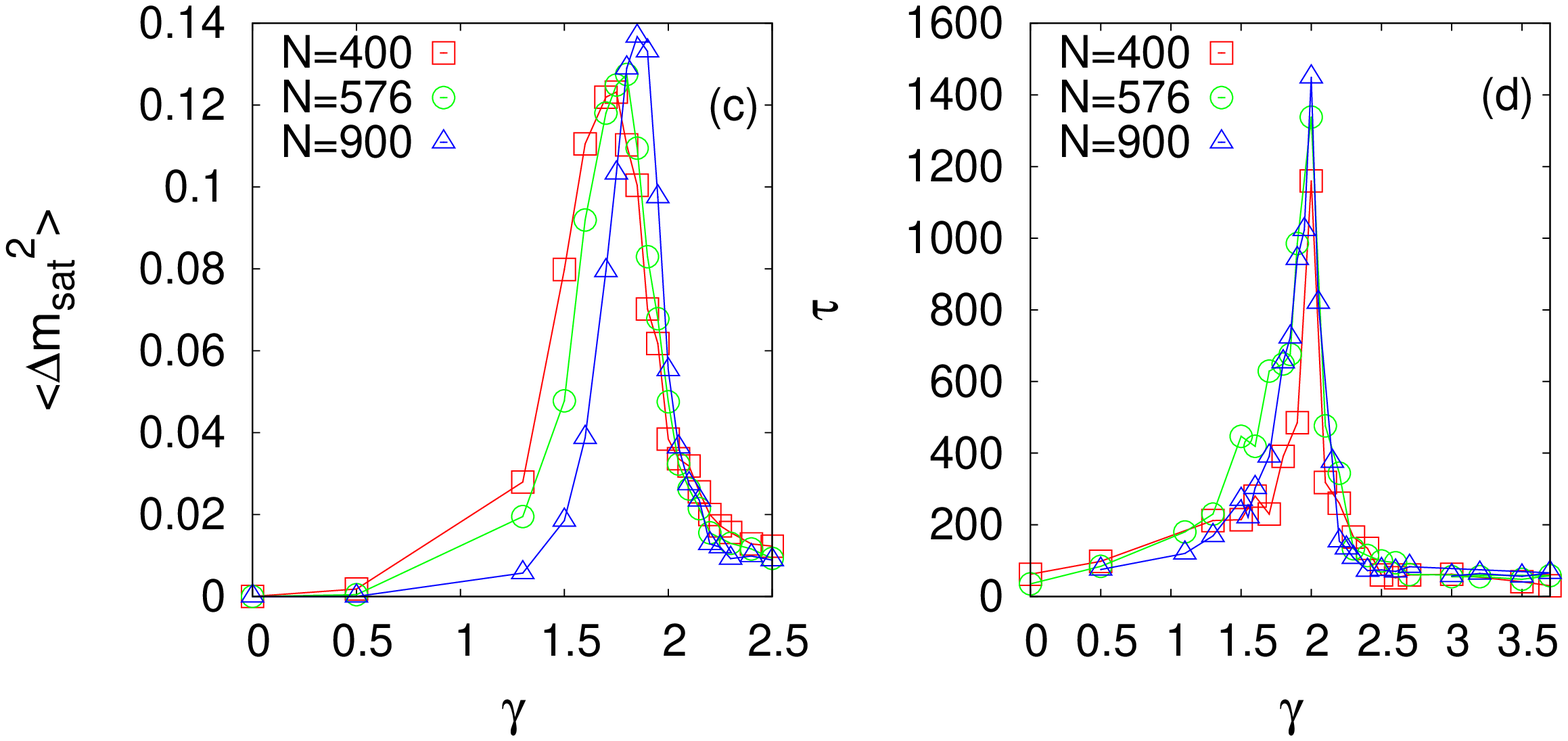}}
\caption{(Color online) RSF: ((a), (b)) The distribution of saturation value of magnetisation for $N=576$ for different values  
of $\gamma$. (c) The variation of fluctuation of saturation value of magnetisation and (d) time to reach the energetically stable state as a
function of $\gamma$ for different system size $N$.
 \label{dist_rsf}
}
\end{figure}

\begin{figure}[!h]
\resizebox{52mm}{!}{\includegraphics {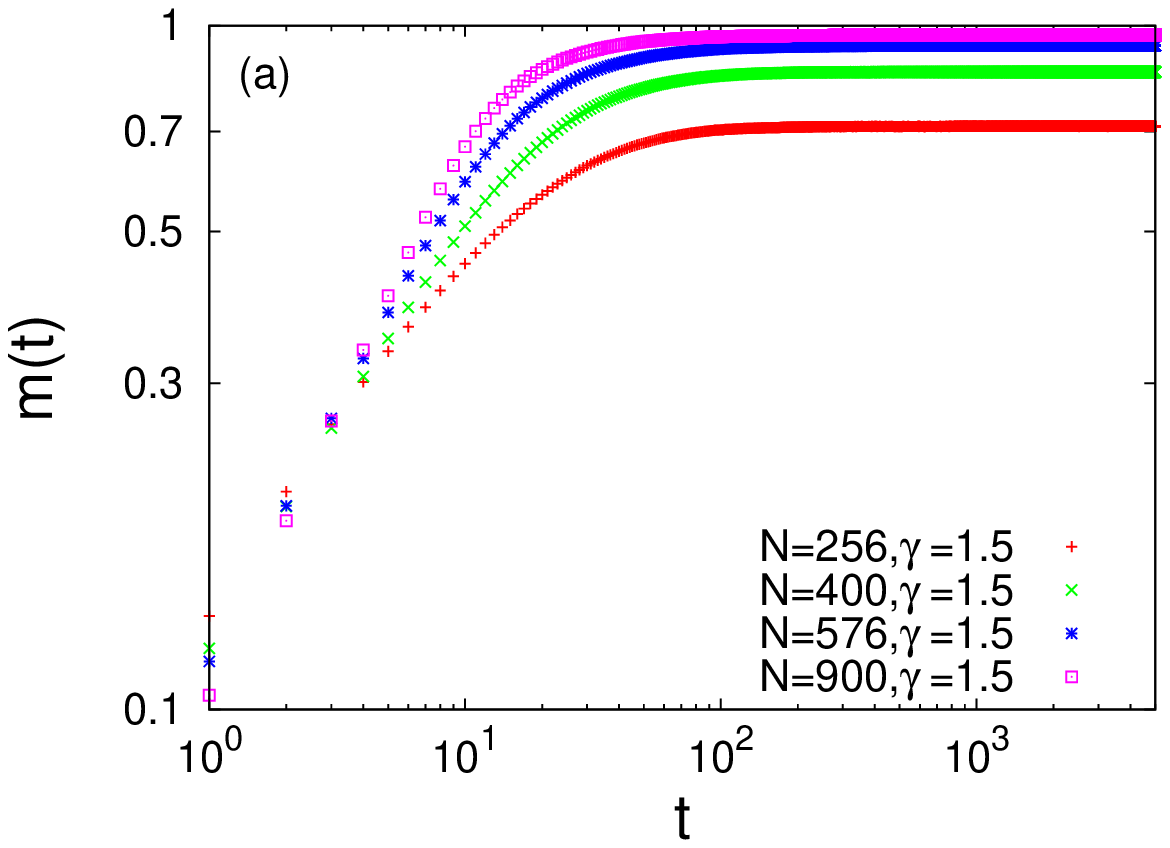}}
\resizebox{52mm}{!}{\includegraphics {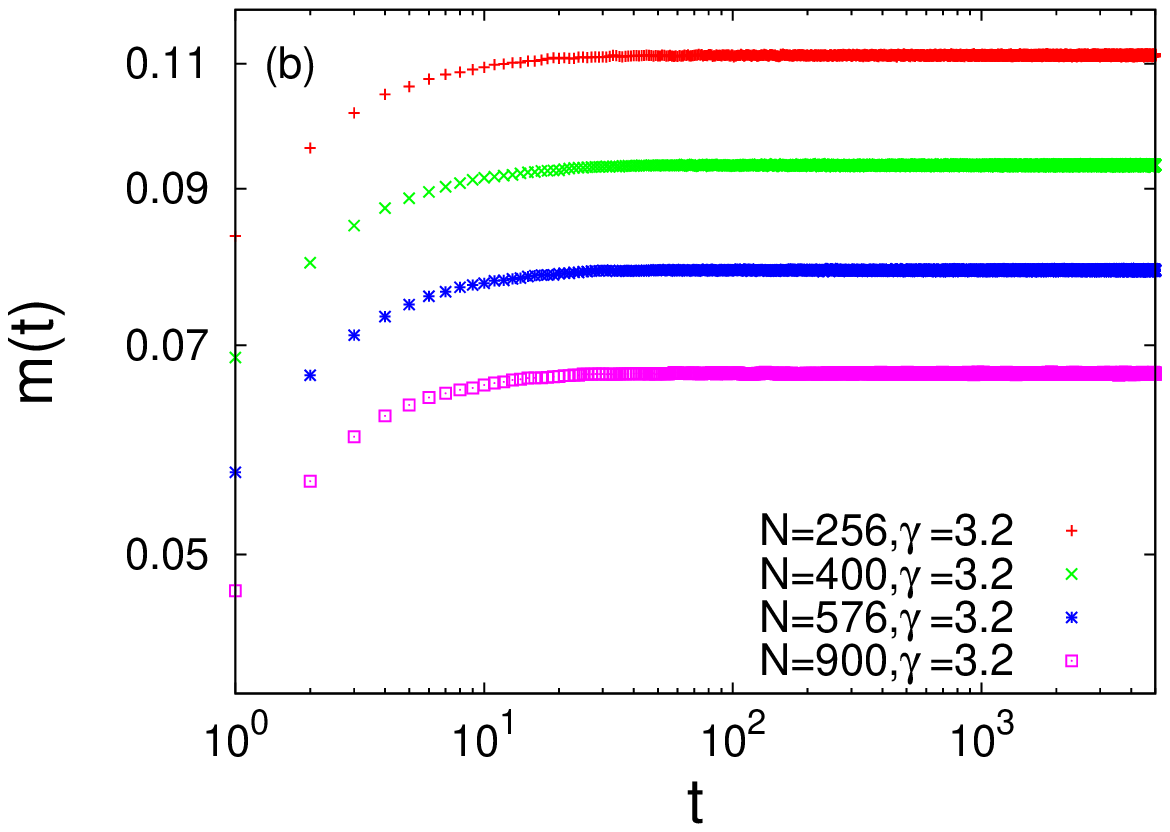}}
\caption{(Color online) RSF: Variation of magnetisation $m(t)$ with time for different system size $N$.
(a) $\gamma=1.5$ and (b) $\gamma=3.2$.
 \label{dyn_mag_rsf}
}
\end{figure}
We first discuss the behaviour of the freezing probability $F$ on the RSF. The variation of $F$
with the network parameter $\gamma$ for different system sizes  
suggests that a freezing transition takes place here as the freezing probability is unity
above  a value of $\gamma$ which we denote by $\gamma_c^f$ (Fig. \ref{sat_frz_rsf}). Of course, if the initial state is fully
ordered it is not counted as a frozen state. However, barring these two states (all up/ all down), none of the other initial states reaches the actual equilibrium configuration.
Hence the freezing probability is actually $1-2/2^N$ which becomes unity in the thermodynamic limit. Below $\gamma_c^f$, 
the freezing probability decreases with system size
and  shows a system size independent behaviour for $\gamma \lesssim 1.0$. The freezing probability is non zero for all values of $\gamma$.

As the freezing probability shows finite size dependence close to $\gamma_c^f$ and 
since it is dimensionless, we argue that it should  show a finite size scaling behavior in the following manner
\begin{equation}
F(\gamma,N)  =   g_1[(\gamma-\gamma_c^f)N^{1/{\tilde\nu}}]
\end{equation}
where $g_1$ is a scaling function. 
 We  estimate  $\gamma_c^f \approx 2.0$ and the exponent $\tilde \nu \approx 2.20$ (inset of Fig.\ref{sat_frz_rsf}). 
 The estimates are obtained from the manifestly best collapse of the data after rescaling is done properly.
As already discussed in section  \ref{ising-net}, 
%\cite{Kim,Herrero,Hong,achat-psen} that on networks, where the dimensionality is ill-defined,
finite size scaling is valid for the Ising model on networks with $\tilde \nu = \nu d$ where $\nu$ is the mean field value of the 
correlation length exponent and $d$ is equal to $4$. 
Assuming the same to hold good here, we get
 $\nu \approx 0.55$ from the fact that $\tilde \nu \approx 2.2$ which is fairly 
 close to the mean field value $0.5$ for Ising model \cite{Stanley}. %(later)

%Although the system shows freezing for the entire parameter 
%space, we show that at $\gamma \simeq 2.0$ an 
% order-disorder phase transition is occurring. 
We next show that an order-disorder phase transition also occurs in this network.
The distribution of magnetisation $\Omega(m_{sat})$  has been studied (Fig. \ref{dist_rsf}a, b). It is unimodal in nature for the values 
of $\gamma$ far from the  $\gamma_c^f$. The peaks occur  at $m_{sat}\sim 1$ for $\gamma \ll \gamma_c^f$ and $m_{sat}\sim 0$ for $\gamma \gg \gamma_c^f$.
The distribution is bimodal close to  $\gamma_c^f$ with the peak values at $m_{sat}\sim 0$ and $m_{sat} \sim 1$.
%In fact for $\gamma <1$, where one observes a finite freezing probability independent
%of system size, the magnetisation is very close to unity even in the frozen phases.
%In contrast, for $\gamma >\gamma_c^f$, the distribution shifts towards 
%$m_{sat}=0$. 

Fluctuations of the saturation value of the magnetisation $\langle {{\Delta {m_{sat}}^2}}\rangle$ shows 
a peak which shifts toward $\gamma_c^f$ and the  the peak value increases 
as the system size is increased (Fig. \ref{dist_rsf}c). We have also estimated the time $\tau$ to reach the energetically stable  state.
It shows a peak close to $\gamma_c^f$. 
Clearly $\tau$    diverges near $\gamma_c^f$,
as the  peak value increases with system size (Fig. \ref{dist_rsf}d).
These behaviour suggest  that there is an order-disorder transition also taking place as $\gamma$ is varied.
 In principle the order-disorder transition may take place at a value $\gamma_c^m \neq \gamma_c^f$
and we make further analysis to estimate $\gamma_c^m$ more accurately. 
We also check whether a mean field-like behaviour is   present for the order-disorder transition.

\begin{figure}[!h]
\includegraphics[width=5cm,angle=270]{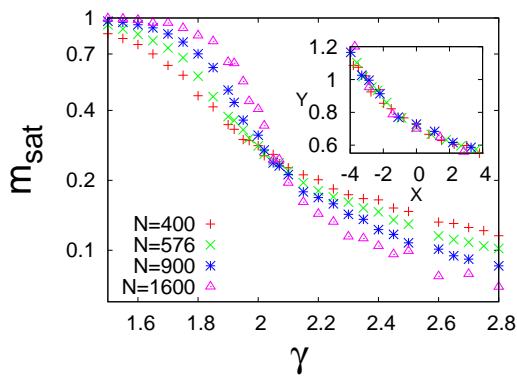}
\caption{(Color online) RSF: Variation of saturation value of magnetisation 
 as a function of $\gamma$ for different system size $N$.
Inset shows the data collapse where $Y=m_{sat}N^{\beta/{\tilde\nu}}$ has been plotted against $X=(\gamma-\gamma_c^m)N^{1/{\tilde\nu}}$.
 \label{sat_mag_rsf}
}
\end{figure}

The variation with time of the magnetisation $m(t)$ has been shown for different system 
sizes  for two different values of the degree exponent $\gamma$ (Fig. \ref{dyn_mag_rsf}a, b).
The behaviour of the saturation values of the magnetisation $m_{sat}$ for finite sizes shows the typical characteristics
of a continuous phase transition with $\gamma$ acting as the driving parameter (Fig. \ref{sat_mag_rsf}).

Using  finite size scaling method we indeed obtain a collapse of the data points  for $m_{sat}$ for different system sizes. 
The following scaling form for $m_{sat}$ has been used:
\begin{equation}
 m_{sat} = N^{-\beta/{\tilde\nu}} g_2[(\gamma-\gamma_c^m)N^{1/{\tilde\nu}}]\\.
\end{equation}
We obtain $\gamma_c^m \approx 2.20$,   
 $\beta \simeq 0.47$ and  $\tilde\nu \simeq 2.20$ (inset of Fig. \ref{sat_mag_rsf}).
The values of the exponents $\beta$ and $\nu$ are fairly 
close to the mean field values once again.

\begin{figure}[!h]
\includegraphics[width=7cm,angle=0]{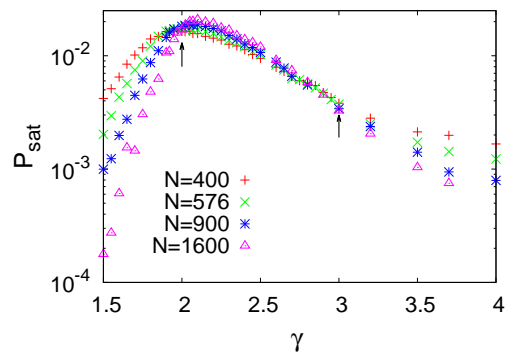}
\caption{(Color online) RSF: Variation of saturation value of the fraction of spin flips 
 as a function of $\gamma$ for different system size $N$. The vertical arrows separate different phases.
 \label{sat_flip_rsf}
}
\end{figure}

\begin{figure}[!h]
\includegraphics[width=7cm,angle=0]{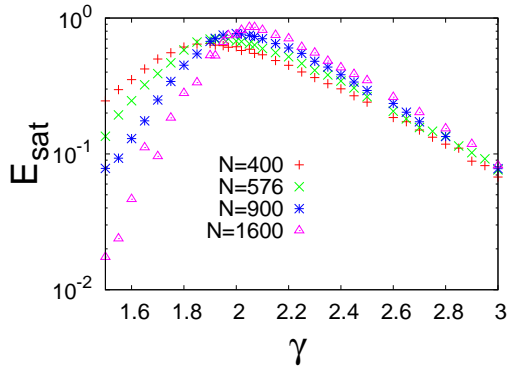}
\caption{(Color online) RSF: Variation of saturation value of the residual energy
 as a function of $\gamma$ for different system size $N$. 
 \label{sat_enr_rsf}
}
\end{figure}

Next we discuss the behaviour of the residual energy and the spin flip probability which 
also attain a saturation value in time ($P_{sat}$ and $E_{sat}$).
Both the saturation values $P_{sat}$ and $E_{sat}$ show nonmonotonic
variation with  $\gamma$ (Figs. \ref{sat_flip_rsf}, \ref{sat_enr_rsf}), with  a peak which shifts  as the system size is increased.
$P_{sat}$ in particular shows a very interesting behaviour with finite size. For $\gamma \lesssim 2.0$, it 
decreases with system size indicating an absorbing phase which is actually the ordered phase as
indicated by the behaviour of the magnetisation discussed above.
However, there is a region between  $\gamma \approx 2.0$ and $\gamma \approx  3.0$, where $P_{sat}$ increases with system size 
which indicates an active state. For $\gamma \gtrsim  3.0$, $P_{sat}$
decreases with system size indicating an absorbing state once again. Hence we have two absorbing phases
separated by an active phase and  two transitions  as the parameter $\gamma$ is varied.
This is reminiscent of two distinct transitions observed in opinion dynamics models \cite{droz,khaleque}.
%One of these absorbing phases  as well as the active phase are disordered phases while the other 
%is an ordered phase. 
%Like $P_{sat}$, $E_{sat}$ also shows a peak which shifts with system size. For $\gamma \lesssim 2.0$, $E_{sat}$ decreases with
%system size which is an ordered  absorbing phase. But for $\gamma \gtrsim 2.0$, $E_{sat}$ increases with system size which is consistent
%with the fact that it is  a disordered state.

 %From all the above observations one may conjecture that the freezing 
%transition and the order-disorder transition actually coincide,
%i.e. $\gamma_c^m=\gamma_c^f=2.0$ in the absence of finite size effects.
We found from the above studies that $\gamma_c^f \approx 2.0$
and $\gamma_c^m\approx 2.2$ are close but not exactly the same. However, this could be due to finite size
effects and  these two might turn out to be
identical, i.e., $\gamma_c^f=\gamma_c^m=2.0$ in the thermodynamic limit. This possibility  is  supported by the behaviour
of both $\tau$ and $E_{sat}$. While $\tau$ diverges at $\gamma \simeq 2.0$, $E_{sat}$  
decreases with system size below $\gamma\simeq 2.0$ (indicating an order phase) and increases or remains
constant above this value (Figs. \ref{sat_enr_rsf}).

%..........................

%%%%%%%%%%%%%%%%%%%%%%%%%%%%%%%%%%%%%%%%%%%%%%%%%%%%%%%%%%%%%%%%%%%%%%%%%%%%%%%%%%%%%%%%%%%%%%%

%Are there 2 transitions - check for $P_{sat}$   ??

\subsection{Barab\'{a}si-Albert network}
\begin{figure}[!h]
\resizebox{42mm}{!}{\includegraphics {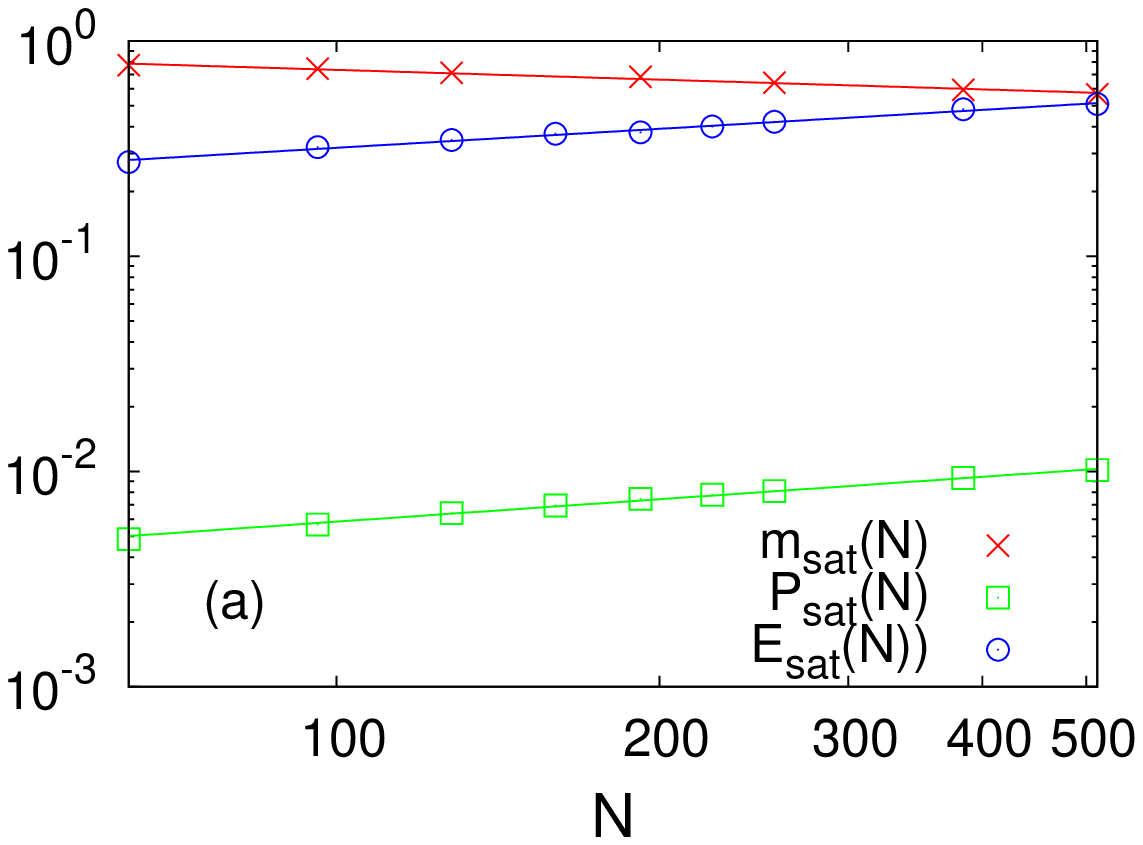}}
\resizebox{43mm}{!}{\includegraphics {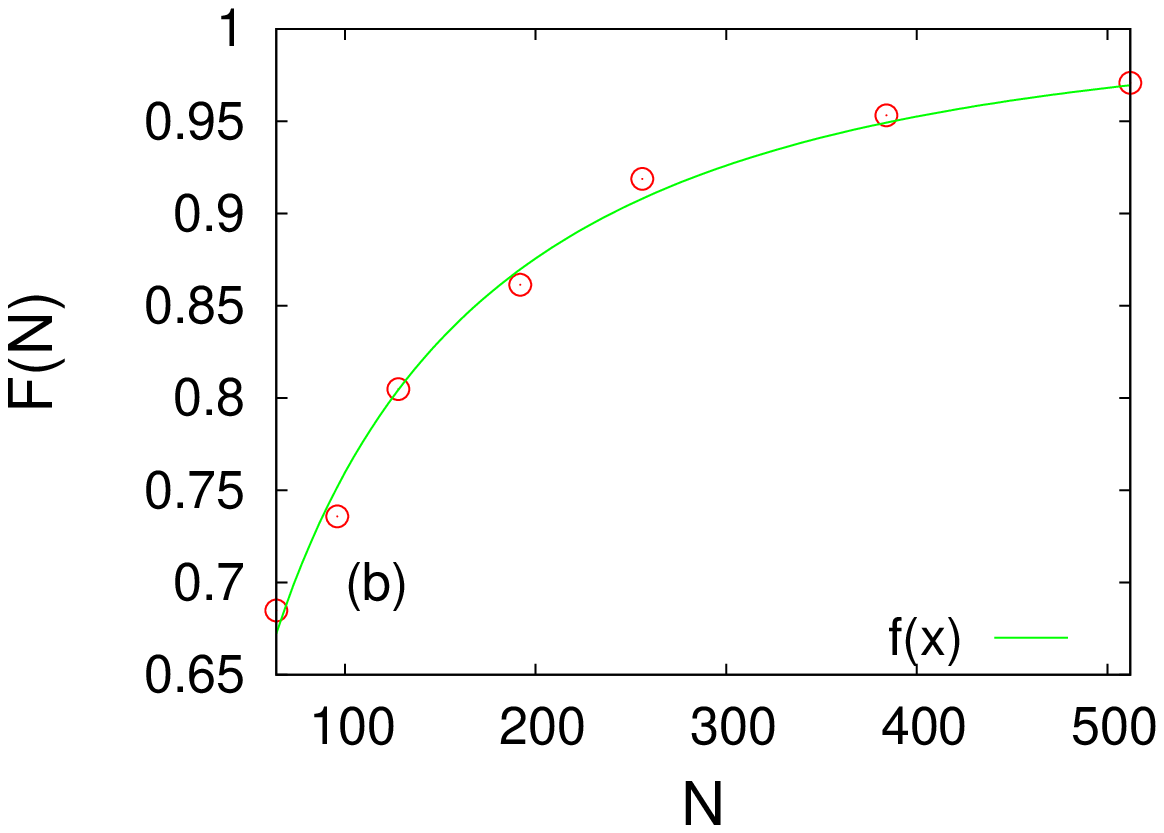}}
\caption{(Color online) BA: (a) Variation of saturation value of magnetisation $m_{sat}(N)$, fraction of spin flips $P_{sat}(N)$ 
and residual energy $E_{sat}(N)$ as a function system size. (b) Variation of freezing probability $F(N)$
 as a function of system size $N$.
 \label{sat_ba}
}
\end{figure}

The BA network has no intrinsic parameter. We calculate the relevant dynamic quantities for different system sizes.
%Like random scale free network, on the  Barab\'{a}si-Albert network we have studied the dynamics of magnetisation ($m(t)$), residual energy ($E_r(t)$) and 
%the fraction of number of spin flips ($P_{flip}(t)$)%(Fig. \ref{dyn_ba})
%for different system sizes.
The magnetisation $m(t)$ slowly increases for the first few time steps and gets saturated at long times for all the system sizes. 
Both $E_r(t)$ and $P_{flip}(t)$ decrease with time and then reach a saturation value. Saturation values of all these quantities
as a function of system size have been plotted in Fig. \ref{sat_ba}a.
The saturation value of magnetisation $m_{sat}(N)$ decreases with system size and the variation shows a power law behaviour.
We fitted the variation with the form $m_{sat}(N)=a_mN^{-b_m}$ and the estimated exponents are $a_m \approx 1.467$ and $ b_m \approx 0.150$.

\begin{figure}[!h]
\includegraphics[width=7cm,angle=0]{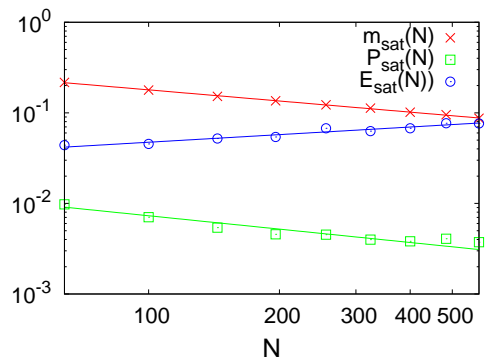}
\caption{(Color online) RSF($\gamma=3.0)$: Variation of saturation value of magnetisation $m_{sat}(N)$, fraction of spin flips $P_{sat}(N)$ 
and residual energy $E_{sat}(N)$ as a function system size.% (left panel). Variation of freezing probability $F(N)$
% as a function of system size $N$ (right panel).
 \label{sat_rsf}
}
\end{figure}

We have also studied the variation of saturation value of the fraction of spin flips $P_{sat}(N)$ and saturation value of
residual energy $E_{sat}(N)$. Both of them show a slowly increasing behaviour with system size and these increasing behaviour  also
follow power law (Fig. \ref{sat_ba}a). We fitted the variation of $P_{sat}(N)$ with the form $P_{sat}(N)=a_fN^{b_f}$
and the exponents are $a_f \approx 0.001$ and $b_f \approx 0.346$. For $E_{sat}(N)$, the fitted form is $E_{sat}(N)=a_EN^{b_E}$ and
the exponents are $a_E \approx 0.083$ and $b_E \approx 0.293$.
The important point to note here is this seems to be an active phase as the 
saturation value of the fraction of spin flips shows increase with system size.
In contrast, the same quantities plotted for the RSF with $\gamma = 3.0$ (Fig. \ref{sat_rsf}) 
shows that it is an absorbing phase (as already noted in the previous subsection).
We will discuss more about this observation in Section \ref{discussion}. 
The variation of $E_{sat}(N)$ with $N$ in RSF is however, similar, with $E_{sat}(N)=0.013N^{0.279}$. $m_{sat}(N)$ for RSF decays as $1.17N^{-0.408}$
which signifies a faster decay compared to the BA network.
.

In the BA  model, the freezing probability $F(N)$ increases with system size (Fig. \ref{sat_ba}b). 
Freezing probability $\rightarrow 1$ for $N \rightarrow \infty$.
Thus the BA model is in an active disordered state where none of the 
configuration reaches the equilibrium ground state.  The variation of
$F(N)$ with system size fits well with the form $F(N)=1-e^{-a_dN^{b_d}}$ and the 
calculated exponents are $a_d \approx 0.113$ and $b_d \approx 0.548$.
The behaviour of the freezing probability as a function of $N$ is also quite 
different for the BA and the RSF networks (with $\gamma = 3$), in the latter we found a system size independent behaviour. 

\subsection{WS network}

On the WS network, all the relevant quantities show a saturation behaviour as already noted previously on
a slightly different version of the WS network \cite{biswas}.

\begin{figure}[!h]
\resizebox{43mm}{!}{\includegraphics {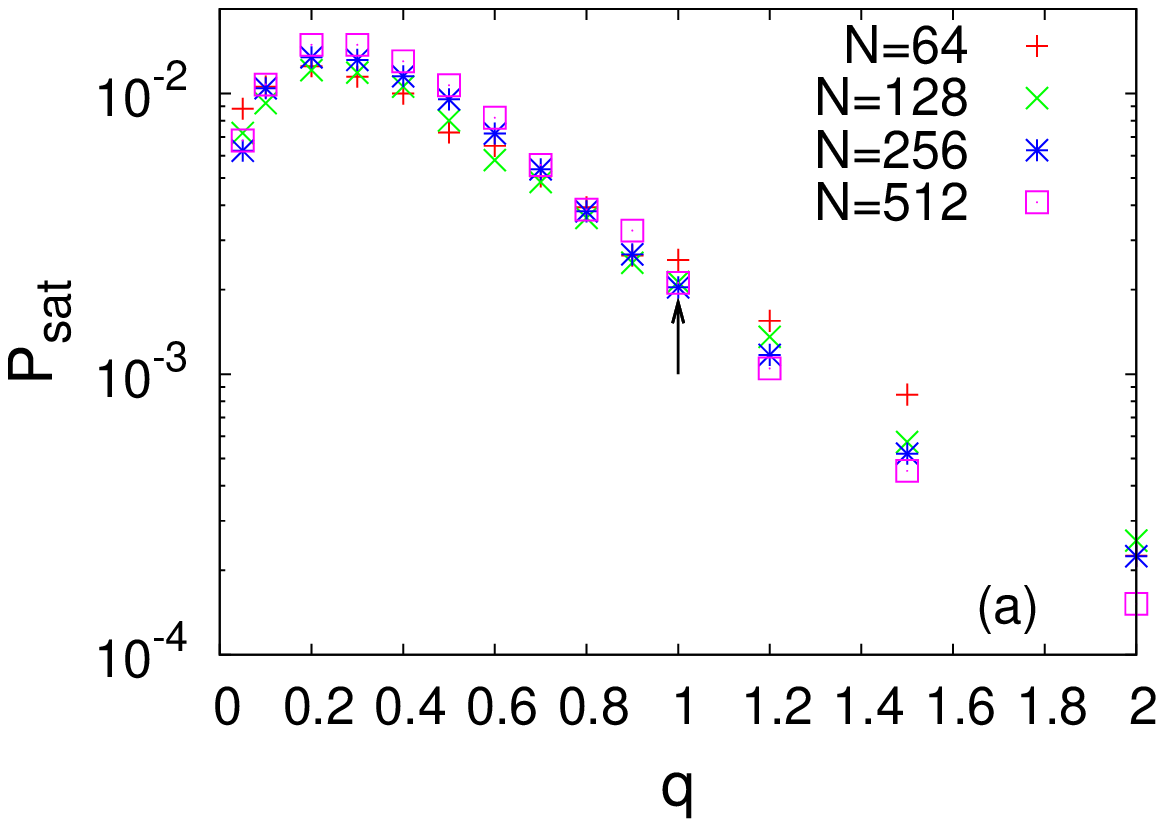}}
\resizebox{42mm}{!}{\includegraphics {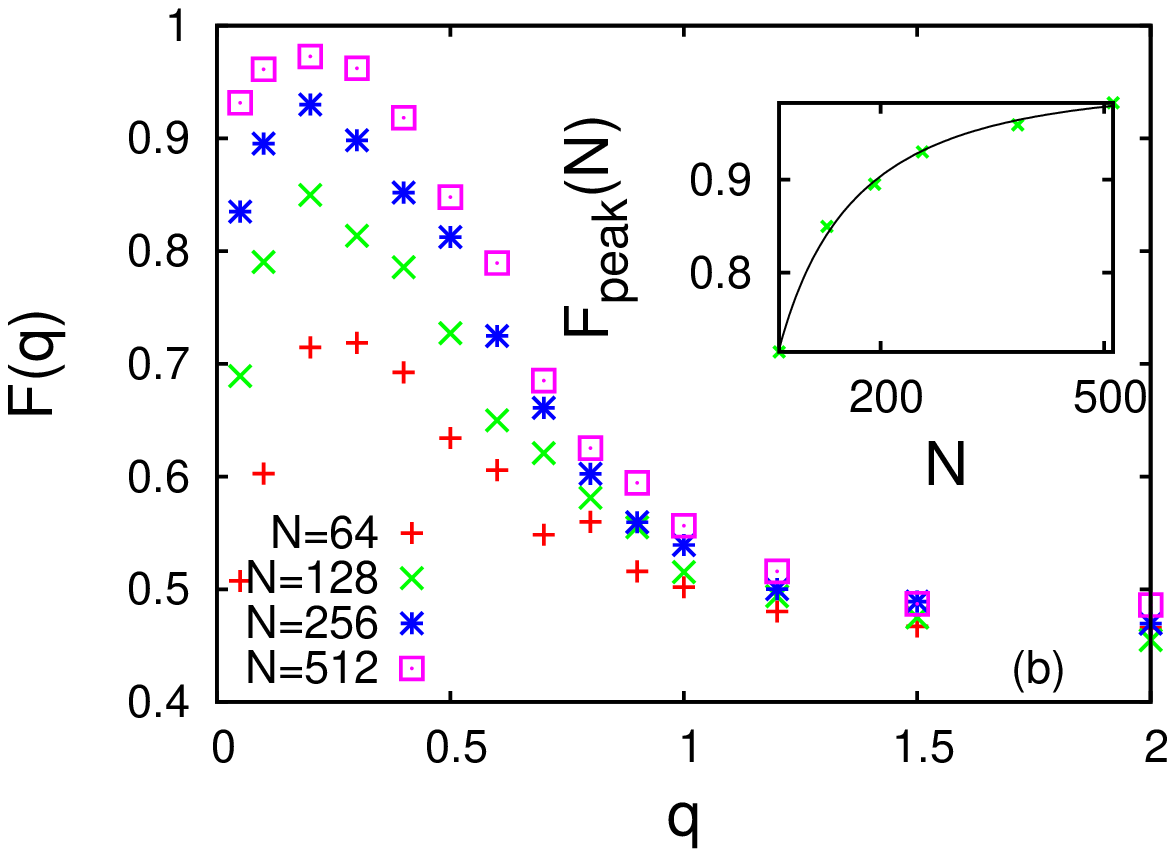}}
\caption{(Color online) WS: Variation of (a) saturation value of fraction of spin flips $P_{sat}$
and (b) freezing probability $F(q)$ as a function of $q$ for different system sizes $N$ (right panel). Inset of (b) shows the variation of peak value of $F(q)$
with the system size.
 \label{sat_ws}
}
\end{figure}

 Here in  addition,  we have  studied the spin flip probabilities. 
%which has not been studied previously for WS like network.
The saturation value of  the
probability of spin flips $P_{sat}$ has been plotted against the parameter $q$ for different system sizes and this quantity reveals interesting behaviour (Fig. \ref{sat_ws}a).
It is clearly seen that above $q \simeq 1$,
the probability decreases as a function of system size $N$ while below $q\simeq 1$ it is
almost size independent. This  indicates that there is an active absorbing phase
transition taking place at this point.

The value of the freezing probability $F(q)$  either increases with $N$ or remains constant which 
clearly shows that the entire phase is frozen for any $q> 0$. We find that   for $q < 1$ the freezing probability reaches unity in the thermodynamic limit 
while it remains fairly constant beyond this value (Fig. \ref{sat_ws}b). This is consistent with the 
active-absorbing phase transition stipulated to take place at $q=1$; in the active state, one can never  reach the equilibrium ground state configuration.

$F(q)$ shows a peak at $q=0.2$  and the position of the peak is independent of system size which
shows that the system is maximally disordered here.
The peak values of freezing probability as a function of system size $N$ is fitted with
the form $F_{peak}(N)=1-e^{-aN^b}$ and the estimated values of the exponents
are $a\approx 0.134$ and $b\approx 0.540$ (inset of Fig. \ref{sat_ws}b). It may be noted that the same form is obeyed in the BA model.

%What about other q values ??

% Scale free networks: 

\section{Discussions and conclusions}
We have studied zero temperature Glauber dynamics of the Ising model on three types of networks and compared the results.
Frozen state is observed in all the three types of network models. For random scale free network, freezing probability is
 unity for $\gamma \gtrsim 2$, i.e., the system never reaches  the global equilibrium
 but for lower value of the parameter $\gamma$, it decreases with system size and shows a system size independent 
behaviour for $\gamma \lesssim 1$. This behaviour of freezing probability suggests a freezing transition point at $\gamma_c^f \simeq 2.0$.
We also find an order disorder transition point taking place very close to this point; in fact we believe that they occur at the same point
and the difference is only a finite size effect. Also close to this point, the first active-absorbing (A-A) phase transition takes place; 
the disordered phase for $2\lesssim \gamma \lesssim 3$ is active while for $\gamma \lesssim 2$, 
one gets an absorbing phase. A second A-A transition takes place close to $\gamma \approx 3.0$
and the system evolves to an absorbing disordered state beyond this value. In all probability these two A-A transitions take place at $\gamma=2.0$ and
$3.0$ in absence of finite size effects; these two points are significant as  for $\gamma\leq 2.0$, the average degree diverges while
for  $\gamma\leq 3.0$, the degree variance diverges in the thermodynamic limit.

One can compare the results of RSF network and BA network for the same characteristic degree exponent $\gamma =3.0$.
The residual energy shows an increase with system size $N$ in both cases in a power law manner with an exponent which is fairly
 close. Also, the saturation value of magnetisation decreases with $N$ in both networks in a power law manner, corresponding exponents are
 however quite different. The saturation energy and magnetisation behaviour are consistent with the fact that the state is disordered for both RSF and BA networks.
  However freezing probability for BA model and RSF network
show different behaviour with system size.  For RSF, the  freezing probability  shows a system size independent behaviour while for  BA model, it has  a nonlinear dependence.
As far as the spin flip probability is concerned RSF (at $\gamma = 3.0$) again differs from the BA network.
The RSF network and BA model are intrinsically different,  BA is a growing network generated with a particular strategy.
There is no loop in this network. In the case of RSF, the structure is completely different, there may be loops. 
Although in a numerical study, the value of $\gamma$ may not be exactly 3 in either the RSF or BA networks
due to finite size effects, we believe that this  cannot be the reason for the results being  
qualitatively different. In fact 
previous studies on RSF and BA have shown that characteristic features may be quite different for the two networks \cite{albert1,psen}.
%However if we consider that the second A-A transition actually takes place at $\gamma = 3.0$, both the networks may be in an active state,
%that this is not observed in the RSF is perhaps due to finite size effect.

The results obtained for the WS model can also be compared to those found for the RSF network and BA network.
In the WS network an A-A transition is observed at $q \simeq 1.0$ which is also the order-disorder transition point. Hence this 
is similar to the occurrence of an A-A and an order-disorder transition occurring simultaneously in the RSF. However in the WS network,
the entire disordered phase is active. For small average degree the system is in an active and for large degree in an absorbing phase.
For the RSF network,  two A-A transitions exist where absorbing phase is observed for {\it both} large degree and small degree and an active 
state exists in between these two absorbing phases. In both RSF and WS another interesting feature is present, the ordered 
state shows a finite freezing probability with negligible system size dependence.
In the WS network maximum value of the freezing probability $F(q)$ occurs at $q\approx 0.2$ and 
shows a behaviour similar to the freezing probability in BA network
as a function of system size.

To summarise, a systematic study of ordering dynamics of the Ising model on scale free networks has been made for the first time to the best of our knowledge.
It is observed that the system freezes to a non equilibrium steady state for all values of the relevant parameters in the random scale free
networks (RSF) and the Barab\'{a}si-Albert model (BA).
The presence of two active-absorbing phase transitions in the RSF makes it different from the WS network where only one such transition 
is observed. It is also concluded that in RSF one of the active-absorbing phase transition takes place at the order-disorder transition
point which is similar to what is observed in the WS network.
\label{discussion}

Acknowledgements: The authors thank Soham Biswas for his suggestions and encouragement.
AK acknowledges financial support from UGC sanction no. F.7-48/2007(BSR). PS acknowledges financial support from CSIR 
project.  Computations made on HP cluster  financed by DST (FIST scheme), India.
{}

\end{document}